# Trapped field in a superconducting disk magnetized with low swept-down rate of applied field

Y Hou, C Y He, L Liu, Z X Gao

Department of Physics, Key Laboratory for Artificial Microstructure and Mesoscopic Physics, Peking University, Beijing 100871, P. R. China

**Abstract:**

We have proposed a systematic theoretic framework to calculate the trapped magnetic field and temperature distributions in a superconducting disk (SD) magnetized by a field-cooling process. Our calculations are based on the critical state model with temperature and field-dependent critical current density, and the heat conduction equation with account of the heat produced by flux motion. We distinguish the normal state region from the superconducting state region. Our calculated results are in good agreement with the recently experimental results reported by different groups, and the deviation is well explained.

**Introduction:**

The recent progress in fabrication of large sized bulk high-temperature-superconductors (HTS) has gained them intense attention for their strong pinning effect and great capability to trap high magnetic fields. The recently reported very high trapped fields [1-3] make HTS promising for use as permanent magnets, which imply further practical applications [4]. Many different potential engineering applications have emerged for such magnets [5-11], such as the magnetic levitation trains, flywheels and magnetic bearings, which have been developed by utilizing the repulsive force against the magnetic fields produced by a permanent magnet.

Three typical ways are often used to magnetically activate HTS: zero-field-cooling magnetization (ZFC), field-cooling magnetization (FC), and pulsed field magnetization (PFM) [12]. FC, which is usually used in the laboratory, is the most effective method to extract the potential of the material. Several experimental approaches have been made to trap very high magnetic fields in superconducting bulk magnets. G. Fuchs *et al.* reported trapped-field of 8.5 T (at 51.5 K) and 14.35 T (at 22.5 K) in a bulk melt textured YBCO sample of 2.6 cm in diameter [2], and S. Gruss *et al.* reported trapped-field of 16 T (at 24 K) and 11.2 T (at 47 K) for a improved sample of the same size [3]. Recently M. Tomita and M. Murakami significantly enhanced the thermal stability and internal mechanical strength of an YBCO bulk of 2.65 cm in diameter, thereby achieved a trapped-field over 17 T at 29 K [1]. All of the very high trapped magnetic fields reported by different groups are obtained at low swept-down rate of the applied magnetic field. Theoretical work has also undergone rapid developments during the past decade. Brandt presented the basic equations and an effective calculation method [13-14], as well as the results of the levitation force between a non-magnetized superconductor and a permanent magnet [15]. Wang *et al.* have reported calculation results of the levitation force between a field-cooling magnetized superconductor and a permanent magnet [16]. But in the above calculations, the magnetic field and temperature dependence of the critical current density, as well as the heat dissipation was not taken into account. S. Bræck *et al.* took both factors into consideration in their calculations concerning a slab-shaped superconductor [17], which stood for a 1-D problem. H. Ohsaki *et al.* also considered both factors, and calculated the distributions of current density and temperature in a pulsed field magnetization process [12]. However, it is not clear whether or not the normal state region is distinguished from the superconducting state region both in the calculations of current density and heat dissipation.

In this paper, we have proposed a systematic theoretic framework to calculate the trapped magnetic field. We not only take into consideration the heat dissipation and temperature and field-dependent critical current density, but also distinguish the normal state region from the superconducting state region by taking different material equations $E(J)$, and we present our calculation of the trapped field and temperature distributions in a SD during a field-cooling magnetization process. Numerical analysis is used in these calculations.

## Equations and Basic Approach:

### A. Modeling:

We consider a SD with radius $a$ and thickness $b$. The superconductor is cooled below its critical temperature in a uniform applied magnetic field. Then the applied magnetic field is swept down to zero at a certain rate. After this field-cooling magnetization process, certain flux is trapped by the superconductor and it behaves as a trapped-filed magnet. We calculate the magnetic field distributions after this magnetization process. The numerical analysis is made on the axisymmetric coordinate ($r$-$z$). Due to the symmetry of the cylindrical system, we make our calculations on half the section plane along the radius, featuring a rectangle of $a$ and $b$.

### B. Basic Equations for Current Density:

Several works have obtained the desired equation of motion for the current density in the cylindrical case as following:

$$\dot{J}(\mathbf{r},t) = \mu_0^{-1} \int_0^a d\mathbf{r}' \int_0^b dz' Q_{cyl}^{-1}(\mathbf{r},\mathbf{r}') \left[ E(J) + \dot{A}_f(\mathbf{r}',z') \right] \qquad (1)$$

Where 
$$Q_{cyl}(\mathbf{r},\mathbf{r}') = f(r, r', z-z') \qquad (2)$$

and $Q^{-1}$ is the reciprocal kernel, which is defined by

$$\int_0^a d\mathbf{r}' \int_0^b dz' Q^{-1}(\mathbf{r},\mathbf{r}') Q(\mathbf{r}',\mathbf{r}'') = \delta(\mathbf{r}-\mathbf{r}'') \qquad (3)$$

With

$$f(\mathbf{r},\mathbf{r}',z-z') = \int_0^P \frac{d\varphi}{2\pi} \frac{-r'\cos\varphi}{\left[(z-z')^2 + r^2 + r'^2 - 2rr'\cos\varphi\right]^{1/2}}$$

$$= \frac{-1}{\pi k}\sqrt{\frac{r'}{r}}\left[\left(1-\frac{1}{2}k^2\right)K(k^2) - E(k^2)\right] \quad \text{and} \quad k^2 = \frac{4rr'}{(r+r')^2 + (z-z')^2}$$

Eq. (1) can be easily time integrated by starting with $J(r,z,t_0) = J_0$ ($J_0$ is the initial current density distribution in the SD) and then by putting $J(r,z,t=t+dt) = J(r,z,t) + \dot{J}(r,z,t)dt$. As soon as the induced current density $J(r,z,t)$ is obtained, the vector potential generated by the induced current density $A_J$ can be derived as,

$$A_J(\mathbf{r},z) = -\mu_0 \int_0^a d\mathbf{r}' \int_0^b dz' \, Q(\mathbf{r},\mathbf{r}') J(\mathbf{r}'). \tag{4}$$

And the radial and axial trapped magnetic field can be written in the form of,

$$B_r = -\frac{\partial A_J}{\partial z}, \quad B_z = \frac{1}{r}\frac{\partial(rA_J)}{\partial r} \tag{5}$$

respectively.

These equations should be supplemented by certain relationships between $J$ and the magnetic field $B$ and the electric field $E$, which depends on the material. In the superconducting state region, where $J < J_c$, $J_c$ is the critical current density, the power-$n$ model is used to describe the nonlinear characteristics of the superconductor [16]:

$$E = E_c \left(\frac{J}{J_c}\right)^n \tag{6a}$$

$n = \sigma - 1$ and $\sigma$ is the flux creep exponent, and in the normal state region, where $J \geq J_c$, we use

$$E(J) = E_c \left(\frac{J}{J_c}\right). \tag{6b}$$

Generally the critical current density depends on both the local field $B$ and the temperature $T$. The Kim model is used to describe the flux density dependence of $J_c$ [12]:

$$J_c = J_{c0} \frac{B_0}{|B| + B_0} \tag{7}$$

where $J_{c0}$ is $J_c$ when $B = 0$ and $B_0$ is a parameter. We include the temperature dependence of $J_c$ as the following equation [12]:

$$J_{c0} = a[1-(\frac{T}{T_{c0}})^2]^2 \tag{8}$$

where $T_{c0}$ is the critical temperature at $B = 0$ and $a$ is constant.

## C. Heat dissipation:

When a superconductor is subjected to a non-stationary external magnetic field, the heat generation rate per unit volume is $W = \vec{E} \cdot \vec{J}$ (9)

The temperature change due to the heat generation is described by the heat diffusion equation.

$$C\frac{\partial T}{\partial t} - k\nabla^2 T = W \tag{10}$$

Here $k$ is the thermal conductivity, and $C$ is the heat capacity per unit volume.

We consider the case of a SD with radius $a$ and thickness $b$. The superconductor is cooled below its critical temperature in certain cooling medium. Thus the boundary temperature keeps stationary, the same as the temperature of cooling medium. The solution to the heat diffusion equation with the constant temperature boundary condition $T(\vec{r},t)|_\Sigma = T_0$ and uniform initial temperature $T(\vec{r},0) = T_0$ can be expressed as:

$$T(\vec{r},t) = T_0 + \int_0^t dt' \iiint_{V'} Q(\vec{r}',t') G(\vec{r},t;\vec{r}',t') d\vec{r}' \tag{11}$$

where $G(\vec{r},t;\vec{r}',t')$ is the Green's function given by:

$$G(\vec{r},t;\vec{r}',t') = \sum_{m=1}^{\infty}\sum_{n=1}^{\infty} \frac{1}{C} \frac{\frac{1}{2\pi} J_0(\frac{\mu_0^m}{a}r')\sin\frac{n\pi}{b}z'}{\frac{a^2}{2}[J_0'(\mu_0^m)]^2 \frac{b}{2}} e^{-\frac{k}{C}[(\frac{n\pi}{b})^2 + (\frac{\mu_0^m}{a})^2](t-t')} h(t-t') J_0(\frac{\mu_0^m}{a}r)\sin\frac{n\pi}{b}z$$

Here $\mu_n^m$ is the $m$th positive root of Bessel function $J_n(x)$.

## D. The calculation method:

From the above discussions, it can be obviously seen that Eq. (1) and the series of Eq. (6), (7),

(8) and (10) are coupled from each other. In order to get the numerical solution, we can deal with these equations in the following steps, which are similar to the procedure reported by S. Braeck *et al*. [17]. First, $J_c$ is taken as the value at the initial temperature and magnetic field, then $J(\vec{r},t)$, $E(\vec{r},t)$ and $B(\vec{r},t)$ with the time evolution can be calculated by Eq. (3). Second, $W(\vec{r},t)$ in Eq. (9) is substituted with the calculated $J(\vec{r},t)$ and $E(\vec{r},t)$. Then the temperature distribution $T(\vec{r},t)$ can be deduced with Eq. (10) and the corresponding Green function expression. In the next step, we can recalculate the temperature and the magnetic field dependent $J_c(\vec{r},t)$ according to Eq. (7) and Eq. (8) with $B(\vec{r},t)$ and $T(\vec{r},t)$ attained in the first step. Finally, we can educe the corrected $J(\vec{r},t)$ and $B(\vec{r},t)$ with new $J_c(\vec{r},t)$ via Eq. (3). We should repeat these procedures until a self-consistent numerical solution is obtained, which means that this solution can be taken as the actual value within a certain tolerance.

## Results and Discussions:

### A. Selection of Parameters

The parameters used in our calculation are taken as follows: $T_{c0} = 92$ K [16]; $C = 0.88 \times 10^6$ J/m$^3$K [17]; $k = 6$ W·m$^{-1}$·K$^{-1}$ [17]; $E_c = 1 \times 10^{-4}$ V/m [12]. All these parameters remain unchanged in the following discussions unless special declaration. Other parameters may vary under different conditions, and will be indicated below.

### The trapped field

First we simulate the experimental results of trapped magnetic field over 14T reported by G. Fuchs *et al*. The trapped magnetic field was measured by Hall sensors sandwiched by two YBCO bulks of 2.6 cm in diameter and 1.2 cm in height. The sample was cooled below its critical temperature to a goal temperature in a magnetic field of 18 T. The applied magnetic field then was swept down at the rate of

0.1 T/min. Therefore we take $a$=1.3 cm and $b$=2.4 cm, $B_a$ = 18T for the applied field and $t$=10800 s for the total magnetization time. As is specified in their paper [2], $J_c(0T,77K) = 3.8 \times 10^8 A/m^2$, and this gives $\alpha = 4.24 \times 10^9 A/m^2$ according to Eq. (8). We take n=21 [16] and $B_0$ = 1T [18] at 22.5 K.

We can calculate the time evolution of the flux density profile in the SD during the field-decreasing process. The total magnetic field includes two parts, one is the applied magnetic field and the other is the induced magnetic field generated by the induced current density in the SD. In the beginning the decrease of the applied magnetic field is small, and the induced magnetic field almost compensates the decrease of the applied magnetic field at the sample center. Thus the total magnetic field at the sample center remains constant, indicating the strong pinning effect. As the applied magnetic field further decreases, the volume, in which the induced current density exists, increases, but the induced magnetic field can not compensate the decrease of the applied magnetic field any more, even at the sample center. After the applied magnetic field is swept down to zero, the induced magnetic field alone contributes to the total magnetic field, which is the trapped magnetic field. The time evolution of the induced magnetic field is shown in Fig. 1 and the total magnetic field shown in Fig. 2. It is observed that at T=22.5 K, the trapped magnetic field is over 14 T at the center of the bulk. At T=51.5 K (77 K), we choose n=10 (5) and $B_0$ = 0.6T (0.5 T) to calculate the trapped magnetic fields again. The calculated trapped magnetic fields and the reported experimental results are shown in Fig. 3. It is obvious that our calculated results are in good agreement with the reported experimental data at the sample center. However, there are no experimental data to be compared with our calculated distribution of the trapped magnetic field along the sample diameter. The swept-down rate of the applied magnetic field plays a crucial role in the trapped magnetic field, the higher rate, the lower trapped magnetic field.

Now we consider simulating the experiment data by M. Tomita and M. Murakami where the trapped magnetic fields over 17T were reported. The trapped magnetic field was measured by Hall sensors sandwiched by two YBCO bulks of 2.65 cm in diameter and 1.5 cm in height. In this case, we take $a$=1.325 cm and $b$=3 cm, $B_a = 17.9T$, $t$=7160 s, and $a = 1.2 \times 10^{10} A/m^2$. We take n=21, 8, 5 and $B_0 = 1T$, 0.6T, 0.5T for T=29K, 46K, 77K respectively. Our calculation results with the reported experimental results are shown in Fig. 4. It is obvious that our calculated results are again in good agreement with the reported experimental data at the sample center.

It is seen in Fig. 4 that, in our calculated distribution of the trapped magnetic field along the sample diameter there is no plateau, which was observed in the reported experimental results at the central region of the SD [1]. We consider that in this sample the Kim model fails and the fishtail effect [2, 19] has to be taken into account. This effect indicates that the critical current density decreases with the increase of the applied magnetic field, and reaches its minimum at the magnetic field range 0.5-1.0 T, then increases with the increase of the applied magnetic field, and reaches its maximum at high magnetic field, which is called peak field. So the Kim model may be used in the low applied magnetic field range, however, in high applied magnetic field range the critical current density according to the fish-tail effect is much higher than that of Kim model. We have not found the experimental data of the critical current density as a function of the magnetic field in reference [1], so it's hard for us to precisely correct our calculation results. If we take into account the fishtail effect, the calculated results and the experimental data will be coincided very well. We are now trying to find the critical current density as a function of the applied magnetic field and discuss the influence of this fishtail effect on trapped magnetic field. The new results will be reported elsewhere.

**B. The temperature**

We do not obtain notable temperature rise in our calculations, which agree well with the experimental reports. The low rate at which the applied magnetic field is swept down is mainly responsible for this. Such low rate allows the heat produced by flux motion to spread to the surroundings before it can induce recognizable temperature rise. If the applied field is swept down in much shorter time, the temperature rise is significant. Fig. 5 shows the calculated temperature rise in the sample at 29 K when the rate is 1.5T/min, while other parameters keep unchanged. This explains why sweeping the applied magnetic field down slow is important in the field-cooling process. Improved thermal conductivity also contributes to prevent temperature rise, which is obvious because it allows heat to dissipate quicker.

## Conclusions:

We have presented our systematic theoretic framework and have calculated the distributions of the trapped magnetic field and the temperature in a SD magnetized by a field-cooling process from first principles. The induced current density in the SD is obtained from the basic current motion equation. The heat dissipation in the SD is taken into account by considering the heat conduction equation and the analytical solution is given in the form of Green Functions. We also distinguish the normal state region from the superconducting state region. Our calculated results are in good agreement with the experimental results reported by different groups. However the calculated distribution of the trapped magnetic field along the sample diameter is deviated from the experimental data, which can be mainly contributed to the fishtail effect.

## Acknowledgments:

This work was supported by the National Science Foundation of China (NSFC 10174004), the Ministry of Science and Technology of China (Project No. NKBRSF-G1999064602) and the TaiZhao Foundation of Peking University.

Figure Captions:

Fig. 1 The time evolution of the induced magnetic field in the sample of 2.6 cm in diameter at T=22K.

Fig. 2 The time evolution of the total magnetic field in the sample of 2.6 cm in diameter at T=22 K.

Fig. 3 The calculated trapped magnetic field in the sample of 2.6 cm in diameter at T=22 K, 52 K, 77 K. The circled dots represent the reported experimental results in reference [2].

Fig. 4 The calculated trapped field in the sample of 2.65 cm in diameter at T=29 K, 46 K, 77 K. The curves with circled dots represent the reported experimental results in reference [1].

Fig. 5 The temperature rise in the sample of 2.65cm in diameter if the applied field is swept down at 1.5 T/min at T=29 K.

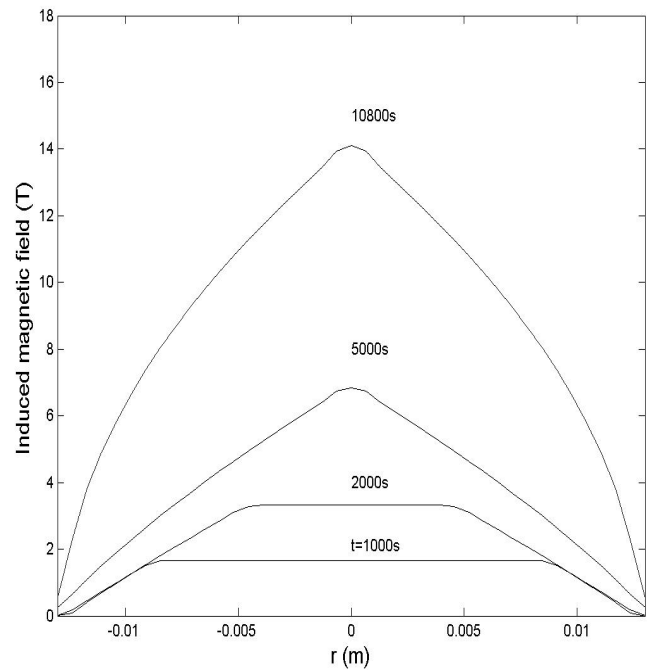

Fig. 1

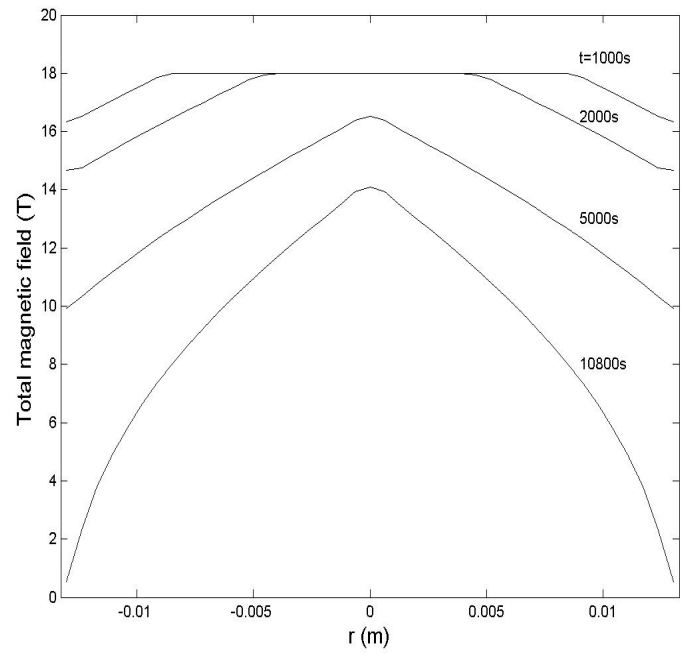

Fig. 2

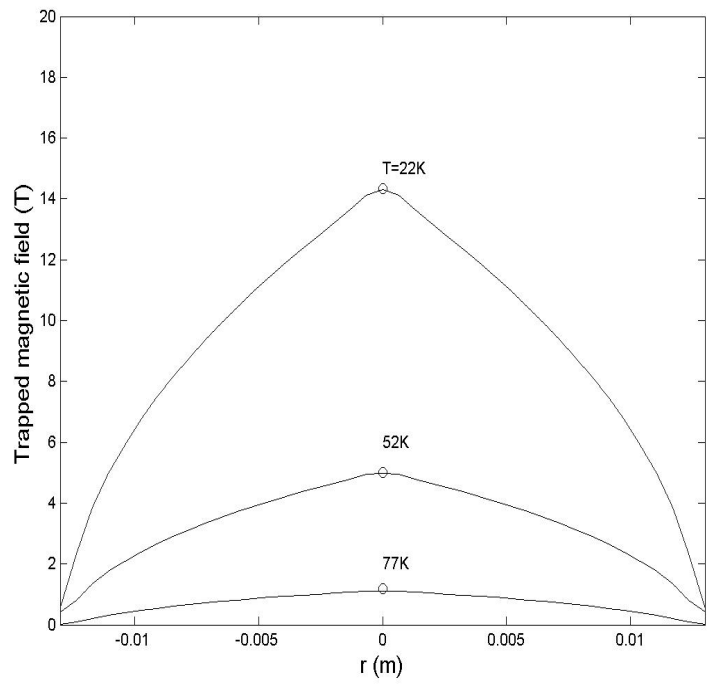

Fig. 3

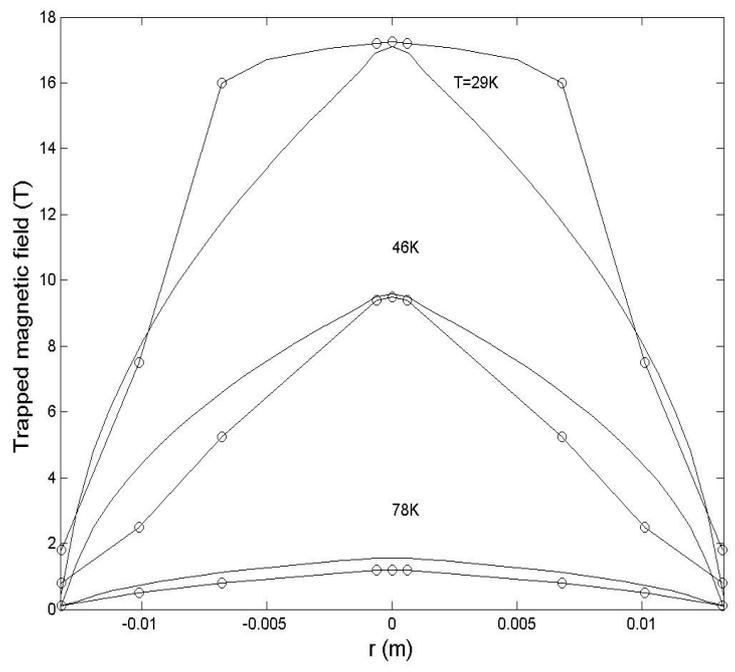

Fig. 4

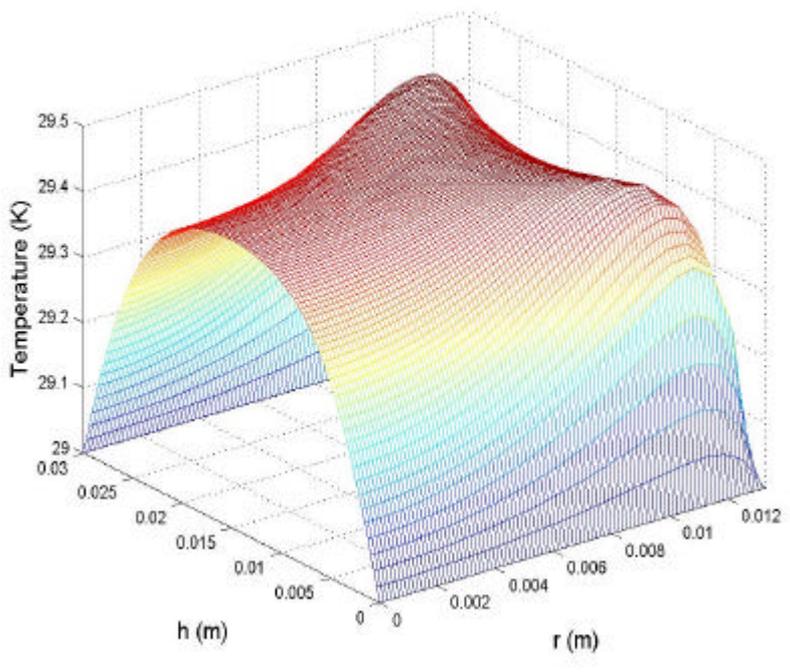

Fig. 5